\documentclass[runningheads]{llncs}
\usepackage[T1]{fontenc}
\usepackage{graphicx}
\usepackage{tabularx}
\begin{document}
\title{MedNuggetizer: Confidence-Based Information Nugget Extraction from Medical Documents\thanks{Preprint accepted at ECIR 2026.}}
\titlerunning{MedNuggetizer}
%
\author{
Gregor Donabauer \inst{1} \and
Samy Ateia \inst{1} \and
Udo Kruschwitz \inst{1} \and
Maximilian Burger \inst{2} \and
Matthias May \inst{3} \and
Christian Gilfrich \inst{3} \and
Maximilian Haas \inst{2} \and
Julio Ruben Rodas Garzaro \inst{3} \and
Christoph Eckl \inst{2}}
\authorrunning{G. Donabauer et al.}
%
\institute{
Information Science, University of Regensburg, Regensburg, Germany \and
Department of Urology, St. Josef Medical Center, Regensburg, Germany \and
Department of Urology, St. Elisabeth Hospital Straubing, Straubing, Germany
\email{\{gregor.donabauer,samy.ateia,udo.kruschwitz\}@ur.de} \\
\email{\{mburger,mhaas,ceckl\}@csj.de} \\
\email{\{matthias.may,christian.gilfrich,julio.rodas-garzaro\}@klinikum-straubing.de}}

\maketitle              
\begin{abstract}
We present \textbf{MedNuggetizer}\footnote{\url{https://mednugget-ai.de/}; access is available upon request.}
, a tool for query-driven extraction and clustering of information nuggets from medical documents to support clinicians in exploring underlying medical evidence. Backed by a large language model (LLM), \textit{MedNuggetizer} performs repeated extractions of information nuggets that are then grouped to generate reliable evidence within and across multiple documents. We demonstrate its utility on the clinical use case of \textit{antibiotic prophylaxis before prostate biopsy} by using major urological guidelines and recent PubMed studies as sources of information. Evaluation by domain experts shows that \textit{MedNuggetizer} provides clinicians and researchers with an efficient way to explore long documents and easily extract reliable, query-focused medical evidence.

\end{abstract}
%
%
%



\section{Introduction}

The increasing adoption of large language models (LLMs) in clinical research \cite{cunha2025medlink,romano2025pie} has introduced new opportunities for automated evidence synthesis, yet it has also raised fundamental concerns about reproducibility. When identical prompts are issued to the same model across repeated runs, variations in outputs can occur due to stochastic sampling, temperature settings, or batch variance \cite{he2025nondeterminism,ouyang2025an,song-etal-2025-good}. Such instability complicates the scientific use of LLM-generated information, particularly when clinical decisions rely on consistent extraction of recommendations.

Our tool \textit{MedNuggetizer} allows medical professionals to assess mitigation strategies and overcome these shortcomings by sampling, clustering, and highlighting extracted information nuggets and the derived confidence of the extraction process. 
Such information nuggets can be used in downstream tasks such as professional search (biomedical systematic reviews) \cite{10.1145/3674127.3674141} or grounded and transparent retrieval augmented generation (RAG) systems \cite{gupta2024overview,Carmel2025LiveRAG}.


By evaluating a use-case on \textit{antibiotic prophylaxis in transrectal and transperineal prostate biopsy} in urology with domain experts, we demonstrate that our tool offers reproducible evidence extraction of recommendations in medicine.

To support the reproducibility of our work, we publicly make available the source code of our app and all data related to our evaluation on GitHub \footnote{\url{https://github.com/SamyAteia/mednuggetizer-ecir2026}}.

\section{System Overview}

We present \textit{MedNuggetizer}\footnote{\url{https://mednugget-ai.de/}; access is available upon request.}
, a user-friendly web application built with Flask. The LLM used in the backend is Google's Gemini 2.5 Flash, chosen for its cost efficiency and its capability of processing PDF content up to 1000 pages \cite{google2025gemini}. The nugget extraction and clustering workflow consists of the following components:


\subsubsection{Information Nugget Extraction}

For each PDF file uploaded by the user, we extract the \textit{information nuggets} relevant to the user’s query that are present within the document. The extraction process is based on AutoNuggetizer \cite{pradeep2024initial}, as implemented in the GINGER framework \cite{lajewska2025ginger}.

\subsubsection{Confidence-Based Clustering}

As variations in LLM outputs can occur even when identical prompts are issued to the same model, we increase the reliability of the extracted information nuggets by providing two parameters: $n$, which defines the number of repeated extraction runs per file, and $conf$, which defines the proportion of the $n$ runs a nugget must appear in to be considered reliable.

To group the extracted nuggets across the $n$ runs, we employ BERTopic-based clustering \cite{Grootendorst2022BERTopicNT}, again following the approach applied in the GINGER framework \cite{lajewska2025ginger}. The minimum number of nuggets required per cluster is determined by $n \times conf$. For example, with $n=5$ and $conf=0.8$, a cluster must contain at least four nuggets to be retained as high-confidence from that PDF file.

\subsubsection{Summary Generation}

After grouping nuggets based on their similarity and recurrence across runs, we aim to consolidate each group of repeated nuggets into a single, high-confidence nugget. For that, an LLM generates a unified nugget that captures the cluster's content in a concise formulation.

\subsubsection{Evidence Guided Information Nugget Clustering}

The input to this final stage consists of all unified nuggets extracted from different files in the previous stage. The objective is to identify semantically similar groups of nuggets across multiple files, as such clusters indicate stronger supporting evidence for the information expressed by these nuggets. To accomplish this, we again apply BERTopic-based clustering \cite{Grootendorst2022BERTopicNT} to the unified nuggets. The clusters identified in this stage form the final output of the \textit{MedNuggetizer} interface. We also provide short headings summarizing identified clusters.

\begin{figure}[ht!]
    \centering
    \includegraphics[width=\linewidth]{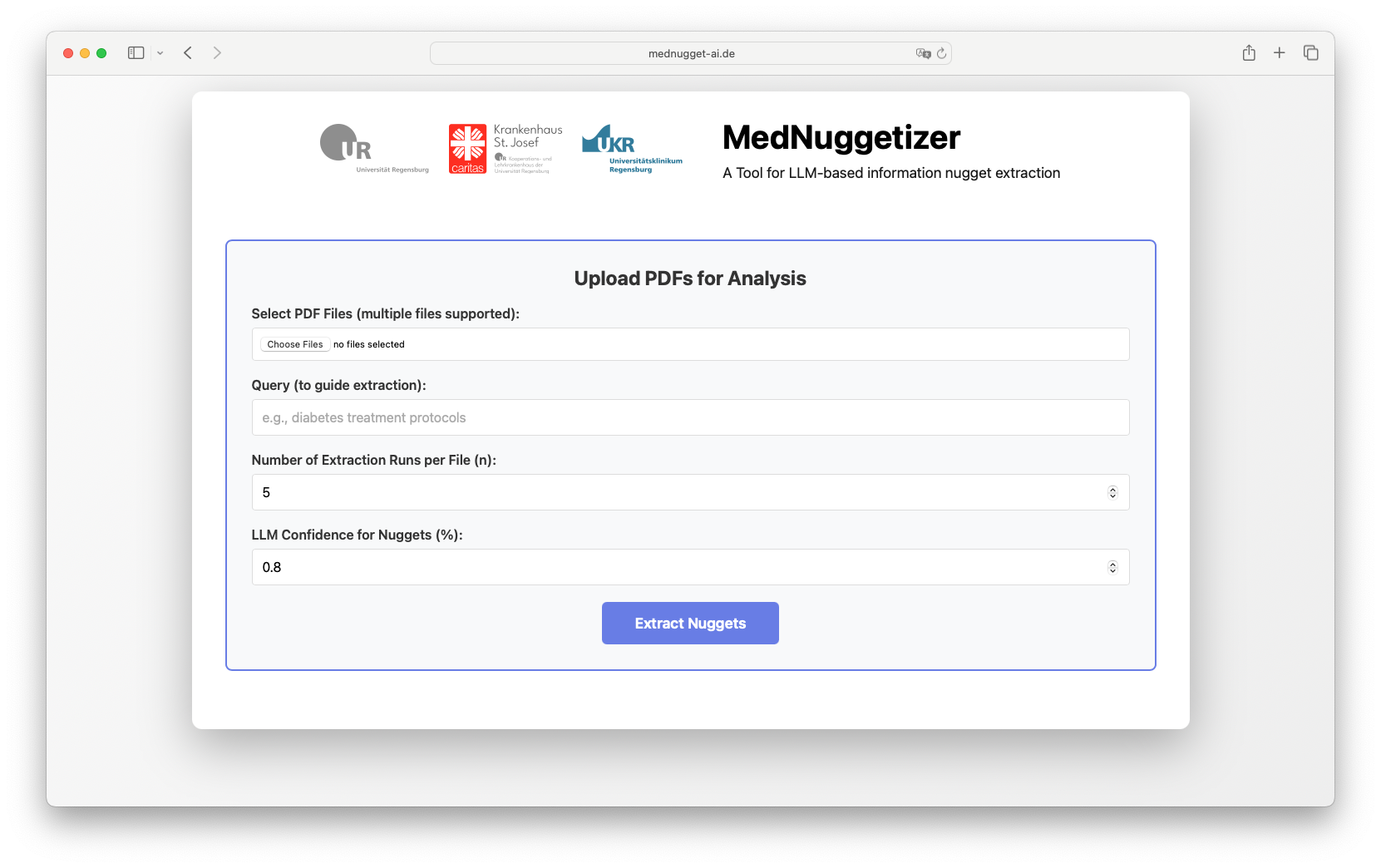}
    \caption{Screenshot of the \textit{MedNuggetizer} web interface. The system allows users to upload PDF files, as well as specifying a query and clustering specific parameters.}
    \label{fig:interface}
\end{figure}

\subsubsection{User Interface}

The interface of \textit{MedNuggetizer} is designed to be self-explanatory and easy to use, as shown in Figure \ref{fig:interface}. It provides four input fields that the user can interact with:

\begin{enumerate}
    \item \textbf{PDF Files:} Allows the user to upload one or more PDF files to be processed.
    \item \textbf{Query:} The query formulated by the user about the uploaded documents; it guides the information nugget extraction process.
    \item \textbf{Number of Runs $n$:} Defines the number of repeated extraction runs performed for each document.
    \item \textbf{LLM Confidence $conf$:} Sets the confidence threshold used to identify reliable clusters of nuggets within the same document.
\end{enumerate}

\section{Evaluation and Discussion}


Following the urologists’ suggestion, we use \textit{antibiotic prophylaxis before prostate biopsy} as an example use case to evaluate \textit{MedNuggetizer}, as existing guidelines and practices vary and optimal preventive strategies remain debated.

We use four major guidelines (EAU Prostate Cancer 2025 \cite{EAU2025_prostate,KRANZ202427}; EAU Urological Infections 2025 \cite{EAU2025_infections}; AWMF S3 Prostate Cancer 2025 \cite{LeitlinienprogrammOnkologie2025}; AWMF S3 Peri-interventional Antibiotic Prophylaxis 2024 \cite{dghm2024s3leitlinie}) and ten recent PubMed-indexed articles (six systematic reviews \cite{TSUBOI202557,STANGL2025,medicina61020198,wolff2024infectious,MADHAVAN2025301,ZATTONI20241303}, four randomized controlled trials \cite{feher2025single,bryant2025local,SADAHIRA202532,doi:10.1177/03915603241273888} published between October 2024 and September 2025) that were identified as relevant by the domain experts. The experts also define five queries\footnote{We provide these as well as the expert assessments on Github.} on the topic which were processed through the tool using the listed PDF documents as context, with hyperparameters set to $n=5$ and $conf=0.8$.

To account for the importance of systematic evaluation in NLP/IR by consulting actual domain experts \cite{eval1,eval2}, two urologists manually annotated the resulting \textbf{(1)} 155 clusters by answering the question \textit{How consistent/coherent is the cluster?}, and \textbf{(2)} 406 information nuggets by answering \textit{How relevant is the nugget to the query?}. Ratings were provided on a five-point Likert scale, where 1 denotes \textit{not at all} and 5 denotes \textit{very}. In addition, they gave general feedback on their experience using the tool.

\begin{table}[h!]
\centering
\caption{Number of clusters/nuggets and their expert-annotated Likert ratings.}
\begin{tabularx}{\textwidth}{|l||X|X|X||X|X|X|}
\hline
\textbf{Query} & $\mathbf{n_{C}}$ & $\mathbf{C_{mean}}$ & $\mathbf{C_{median}}$ & $\mathbf{n_{N}}$ & $\mathbf{N_{mean}}$ & $\mathbf{N_{median}}$ \\
\hline
q0 & 28 & 4.0 & 4 & 66 & 4.0 & 4 \\
q1 & 34 & 4.24 & 4 & 97 & 4.23 & 4 \\
q2 & 44 & 4.77 & 5 & 103 & 4.64 & 5 \\
q3 & 25 & 4.84 & 5 & 65 & 4.75 & 5 \\
q4 & 24 & 4.75 & 5 & 75 & 4.72 & 5 \\
\hline
\end{tabularx}
\label{table:results}
\end{table}

The numbers in Table \ref{table:results} highlight both the high relevance of the information nuggets and the usefulness of the clusters in organizing that information. This aligns with the overall feedback of the urologists, noting that the system can retrieve highly relevant information nuggets, with clusters effectively distinguishing multiple layers of information such as context, current evidence, recommendations, expert opinions, and research limitations. However, identified issues with the nuggets are: undefined abbreviations, missing contextualization, inclusion of method-focused information with limited utility, and partial overlap between clusters, which can hinder clarity and practical applicability.

\section{Conclusion}

We present a tool that allows clinicians to explore medical evidence aggregated from multiple documents, without requiring expertise in underlying IR techniques adopted from SOTA frameworks based on information nugget extraction such as GINGER \cite{lajewska2025ginger}. Its usefulness is demonstrated through a urology use case. Future work could allow users to select backend methods, such as different clustering techniques or the LLMs driving the extraction process.

%
%
%
\bibliographystyle{splncs04}
\bibliography{mybibliography}

\end{document}